\documentclass[12pt,amsmath,amssymb]{article}
\usepackage{amsfonts,amsmath,amssymb}
\usepackage{hyperref}
\usepackage{amsmath}
 
 \newcommand{\bea}{\begin{eqnarray}}
\newcommand{\eea}{\end{eqnarray}}

\newcommand{\pa}[1]{\left(#1 \right)}

 \def\frac#1#2{{#1\over #2}}

 \def\s{\sqrt}

\def\be{\begin{equation}}
\def\ee{\end{equation}}
\def\ba{\begin{eqnarray}}
\def\ea{\end{eqnarray}}

 \def\de{\partial}

 \def\lr{\leftrightarrow}
 \def\f {\frac}
 \def\ti{\tilde}
 \def\ap{\alpha}

 \def\ddd{\cdot\cdot\cdot}
 \def\no{\nonumber \\}

 \def\la{\langle}
 \def\lb{\rangle}

 \def\ov{\overline}

\usepackage{color}
\usepackage{graphicx}
\usepackage{dcolumn}
\usepackage{bm}

\begin{document}

\begin{titlepage}
\thispagestyle{empty}

\begin{flushright}
YITP-18-37
\\
IPMU18-0072
\\
\end{flushright}

\bigskip

\begin{center}
\noindent{{\Large ETH and Modular Invariance of 2D CFTs}}\\
\vspace{2cm}
Yasuaki Hikida$^a$,
Yuya Kusuki$^a$ and
 Tadashi Takayanagi$^{a,b}$
\vspace{1cm}
\\
{\it
$^{a}$Center for Gravitational Physics, \\
Yukawa Institute for Theoretical Physics (YITP), Kyoto University, \\
Kitashirakawa Oiwakecho, Sakyo-ku, Kyoto 606-8502, Japan\\ \vspace{3mm}
$^{b}$Kavli Institute for the Physics and Mathematics of the Universe,\\
University of Tokyo, Kashiwano-ha, Kashiwa, Chiba 277-8582, Japan\\
}
\vskip 2em
\end{center}

\begin{abstract}
We study properties of heavy-light-heavy three-point functions in two-dimensional CFTs
by using the modular invariance of two-point functions on a torus.
We show that the result is non-trivially consistent with the condition of ETH (Eigenstate Thermalization Hypothesis). We also study the open-closed duality of cylinder amplitudes and derive behaviors of disk one-point functions.
\end{abstract}

\end{titlepage}


\section{Introduction}

Two-dimensional conformal field theories (2D CFTs) have provided us an ideal factory of new insights on dynamical properties in quantum field theories. This is highly owing to the strong
constraints imposed by their infinite dimensional conformal symmetries \cite{BPZ}. One well-known
highlight is the Cardy formula
\be
D(E)\sim e^{4\pi\s{\frac{cE}{12}}}  \label{cfla},
\ee
which offers a universal formula for the degeneracy $D(E)$ of highly excited states (with energy $E\gg c$) for general unitary 2D CFTs with the central charge $c$ \cite{CardyF}. This formula is derived from the modular invariance of torus amplitude. In the presence of boundaries, we can employ the open-closed duality to constrain the behaviors of boundary states \cite{Ishibashi:1988kg} in boundary conformal field theories \cite{Cardy:1989ir}. Moreover, the conformal bootstrap for correlation functions leads to another strong constraints on the properties of CFTs, which has successfully been applied not only in two-dimensions but also in higher-dimensions
\cite{Rychkov,SD}.

Recently, there have been very interesting progresses on the properties of three-point functions in 2D CFTs. The modular invariance of torus one-point function leads to a universal formula for the
diagonal part of heavy-light-heavy three-point functions \cite{KM}. The genus two modular invariance gives another constraints on the behavior of heavy-heavy-heavy three-point functions \cite{CMM}, see also \cite{Keller:2017iql,Cho:2017fzo}. The behaviors of
heavy-light-light three-point functions have been worked out in \cite{Das:2017cnv,Ku}.  Refer to
\cite{Hellerman:2009bu,Friedan:2013cba,Hartman:2014oaa,Chang:2015qfa,
Chang:2016ftb,Collier:2016cls,Collier:2017shs,Cho:2017oxl} for other aspects of
recent developments on constraints in 2D CFTs.

Motivated by these developments, the purpose of this article is to explore more of such universal properties through the modular invariance of two-point functions on a torus \cite{Moore:1988uz,Sonoda:1988fq,KSS} and the open-closed duality \cite{Cardy:1989ir}. As we will explain later, the former analysis leads to interesting constraints for off-diagonal part of  heavy-light-heavy three-point functions.
The results satisfy the condition of ETH (Eigenstate Thermalization Hypothesis) \cite{ETH,ETHR,ETHRR}, which is a well-known criterion for a closed quantum system to become chaotic such that it gets thermalized. We will also generalize this argument to multi-point functions on a torus. The latter analysis provides new universal properties on the one-point functions on a disk or equally the coefficients of boundary states. Our analysis below focuses only on the exponential contributions neglecting the polynomial factor of the energy of relevant states, which is denoted by the symbol $\sim$. Refer to \cite{FKW,Lashkari:2016vgj,HLZ,BDDP,LDL,FW} for earlier arguments on ETH in CFTs.

This article is organized as follows:
In section two, we study the modular invariance of two-point functions on a torus and derive a property of three-point functions. After briefly introducing the ETH,
we show that the result satisfies the condition of ETH.
In section three, we study the open-closed duality and derive constraints on
the coefficients of boundary states. We also give a holographic interpretation.
In section four, we summarize our conclusions and discuss
future problems.

{\it Note Added:}
When we were writing up this article, we noticed the interesting preprints \cite{Brehm:2018ipf,BGS} on the arXiv, which have substantial overlaps with the section two of our present one.

\section{Modular Invariance on Torus and ETH}

Consider a two-point function $\la O(t,\phi)O(0,0)\lb$ of a primary operator $O$ on a torus in a given 2D CFT. We choose $O$ such that its spin is vanishing. Its (chiral) conformal dimensions are given by $\Delta_O=\bar{\Delta}_O$ and its energy reads $E_O=2\Delta_O$. The Euclidean time coordinate $t$ and space coordinate $\phi$ are periodic such that $t\sim t+\beta$ and $\phi\sim \phi+2\pi$ on this torus.

We would like to study the constraint imposed by the modular invariance (where we follow the convention in \cite{KSS}). By setting $\phi=0$ for simplicity, the modular invariant relation $\beta\lr (2\pi)^2/\beta$ is written as
\ba
&& e^{\frac{\beta c}{12}}\sum_A\sum_B|\la A|O|B\lb|^2 e^{-(\beta-t)E_A-tE_B}  \no
&& =e^{\frac{(2\pi)^2 c}{12\beta}}\cdot \left(\frac{2\pi}{\beta}\right)^{2E_O}
\sum_A\sum_B e^{-\frac{(2\pi)^2E_A}{\beta}}e^{\frac{2\pi iJ_{AB}t}{\beta}}
|\la A|O|B\lb|^2,  \label{minv}
\ea
where $A$ and $B$ label all states in the CFT including both primaries and descendants;
we set $E_A=\Delta_A+\bar{\Delta}_A$ as the energy of the state $|A\lb$ and $J_{AB}$ as the difference of the spins of $|A\lb$ and $|B\lb$. The matrix element $\la A|O| B\lb$ is equal to the three-point
function $\la A(\infty)O(1)B(0)\lb$.

Below we would like to study implications of (\ref{minv})
in the high temperature limit $\beta\to 0$ with $t /\beta$ kept finite and non-vanishing.
In this case, we do not need to worry about the divergence due to the coincidence of two operators (i.e., $t=0$). In this limit we can set $A$ to be the vacuum.
By using the Cardy formula (\ref{cfla}) for the degeneracy of high energy states $E_{A,B}\gg c$, we can estimate the both sides as follows
\ba
\int dE_A dE_B e^{4\pi \s{\frac{c}{12}E_A}}e^{4\pi \s{\frac{c}{12} E_B}}
\overline{|\la A|O|B\lb|^2} e^{-(\beta-t) E_A-tE_B} \sim e^{\frac{\pi^2 c}{3\beta}}\cdot \left(\frac{2\pi}{\beta}\right)^{2\Delta_O}.  \label{ocxxxx}
\ea
Here, the average is over all states $A$ and $B$ of fixed dimensions $E_A$ and $E_B$.
Since in general $E_A$ and $E_B$ can be arbitrary large in the limit $\beta\to 0$, the above relation
(\ref{ocxxxx}) provides a constraint on the mean squared of heavy-light-heavy three-point functions $\overline{|\la A|O|B\lb|^2}$
as a function of $E_A$ and $E_B$. In the analysis below we only focus on the exponential contributions neglecting the polynomial factor of the energy $E_A$ and $E_B$.

When $t/\beta$ is very small, the dominant contributions of the left hand side of (\ref{ocxxxx})
come from the states with $E_B\gg E_A$. Also if we strictly set $t \to 0$, then the two-point function should have a power
divergence $\sim t^{-2E_O}$. Thus we find that the summation over $B$ should marginally converge when $t/\beta$ is very small, and this leads to the following estimation of the exponential suppression of the three-point functions when
$E_B\gg E_A$:
\ba
\overline{|\la A|O|B\lb_{off}|^2} \sim e^{-4\pi\s{\frac{c}{12}E_B }}.  \label{asymth}
\ea

We would also like to comment that if we keep $\beta$ finite and take the limit $t\to 0$, then the two-point function
behaves as $\la O(t,0)O(0,0)\lb\simeq t^{-2E_O}\cdot Z(\beta)$, where $Z(\beta)$ is the vacuum partition function.
This relation is equivalent to the conformal bootstrap constraint for four-point functions studied in \cite{Das:2017cnv,Ku}.

\subsection{Constraints on 3pt Functions}\label{secthp}

 Now we would like to estimate the square of off-diagonal part of the three-point function $\overline{|\la A|O|B\lb_{off}|^2}$ from the modular invariance (\ref{ocxxxx}).

Let us introduce the parameter $s$ instead of $t$:
\ba
\beta-t=\frac{\beta+s}{2},\ \ \ \ t=\frac{\beta-s}{2},
\ea
where $-\beta\leq s\leq \beta$. Moreover, we parameterize $E_A$ and $E_B$ by ``boost coordinate''
 $(\rho,\theta)$ as follows
\ba
E_A=\rho e^\theta,\ \ \ E_B=\rho e^{-\theta},
\ea
where $-\infty <\theta<\infty$ and $0<\rho<\infty$.

We can naturally assume the following behavior of the three-point function square:
\ba
\overline{|\la A|O|B\lb_{off}|^2} \sim e^{-4\pi\s{\frac{c\rho}{12}}f(\theta)},
\ea
where $f(\theta)$ is an unknown function, which we want to determine from the modular property
(\ref{ocxxxx}) below.

The left hand side of (\ref{ocxxxx}) is estimated by the integral:
\ba
&& \int \rho d\rho d\theta e^{I(\rho,~\theta)}, \no
&& I(\rho,\theta)\equiv 4\pi\s{\frac{cE_A}{12}}+4\pi\s{\frac{cE_B}{12}}-4\pi\s{\frac{c\rho}{12}}f(\theta)
-(\beta-t)E_A-tE_B \no
&& \ \ \ \ \ \  \ =8\pi \s{\frac{c\rho}{12}}\cosh\left(\frac{\theta}{2}\right)
-4\pi\s{\frac{c\rho}{12}}f(\theta)-\beta\rho\cosh\theta-s\rho\sinh\theta.
\ea

Now we apply the saddle point approximation with respect to the integral of $\rho$.
This leads to the following relation by solving $\de_\rho I=0$:
\ba
\s{\rho}=2\pi\s{\frac{c}{12}}
\left(\frac{2\cosh\left(\frac{\theta}{2}\right)-f(\theta)}{\beta\cosh\theta+s\sinh\theta}\right).
\label{wwe}
\ea
By substituting this and assuming $f(\theta)<2\cosh\left(\frac{\theta}{2}\right)$, $I(\rho,\theta)$ is simplified:
\ba
I(\rho,\theta) = \frac{\pi^2 c}{3}\cdot \frac{\left(2\cosh\left(\frac{\theta}{2}\right)-f(\theta)\right)^2}
{\beta\cosh\theta+s\sinh\theta}.
\label{uuuu}
\ea
Note that if $f(\theta)>2\cosh\left(\frac{\theta}{2}\right)$,
then there is no saddle point (\ref{wwe}) and  the related contributions can be neglected.

Now we would like to impose the modular invariance (\ref{ocxxxx}).
This means the following equivalence
in the limit $\beta\to 0$ and $s/\beta=$finite:

\ba
\int^\infty_{-\infty}d\theta e^{\frac{\pi^2 c}{3}\cdot \frac{\left(2\cosh\left(\frac{\theta}{2}\right)-f(\theta)\right)^2}
{\beta\cosh\theta+s\sinh\theta} }\sim e^{ \frac{\pi^2 c}{3\beta}}. \label{gggq}
\ea
This approximated equality should be true for any values of the ratio $s/\beta$ (remember
$s$ takes the values such that $|s/\beta|\leq 1$).
The important point is that the right hand side does not depend on the parameter $s$ and this is possible
only if the contribution from the integral of $\theta$ is localized around $\theta=0$.
This requires (i) $f(0)\simeq 1$ and  (ii) the integral for $\theta\neq 0$ can be negligible, which is equal to the inequality:
\ba
\frac{\left(2\cosh\left(\frac{\theta}{2}\right)-f(\theta)\right)^2}
{\beta\cosh\theta+s\sinh\theta}<\frac{1}{\beta},  \label{wwr}
\ea
for any $|s|\leq \beta$. This inequality (\ref{wwr}) is equivalent to
\ba
f(\theta)\geq 2\cosh\left(\frac{\theta}{2}\right)-e^{-\frac{|\theta|}{2}}=e^{\frac{|\theta|}{2}}.
\ea
where we used $f(\theta)<2\cosh\left(\frac{\theta}{2}\right)$.
Moreover, the previous argument which derives (\ref{asymth}), tells us $f(\theta)\simeq e^{\frac{|\theta|}{2}}$ in the limit $|\theta|\to \infty$.

In summary, we find the following behavior:%

\ba
&& {\mbox{If}}\ \  E_A\simeq E_B\gg c, \ \ \ \overline{|\la A|O|B\lb_{off}|^2}\sim e^{-\f{S(E_A)+S(E_B)}{2}}, \label{ethfa}\\
&&  {\mbox{If}}\ \  c\ll E_A\ll E_B, \ \ \ \overline{|\la A|O|B\lb_{off}|^2} \sim e^{-S(E_B)},  \label{eeeq} \\
&&  {\mbox{For generic values of $E_{A,B}\gg c$}}, \ \ \ \overline{|\la A|O|B\lb_{off}|^2} \lesssim \mbox{Min}
\left[e^{-S(E_A)},e^{-S(E_B)}\right],  \no \label{ethfb}
\ea
where $S(E)=4\pi\s{\frac{cE}{12}}$ is the entropy for states with the energy $E$.

In the above analysis, we ignored contributions from the diagonal part
$\la A|O|B\lb_{diag}=\delta_{AB}\cdot \la A|O|A\lb$. The averaged diagonal three-point function is found in \cite{KM}
\be
\overline{\la A|O|A\lb} \sim e^{-\frac{\pi c}{3} \pa{1-\s{1-12 \frac{E_\chi}{c}}}\s{\frac{12E_A}{c}}}, \label{diagt}
\ee
where $E_\chi$ is the energy of the lowest dimensional state $|\chi\lb$ which satisfies $\la \chi|O|\chi\lb\neq 0$.
 By explicitly substituting this to (\ref{ocxxxx}), we find that the diagonal ones are not dominant over the off-diagonal ones and this justifies the above analysis.

Note that actually we can show that in the limit $E_B \to \infty$ with $E_B-E_A$ fixed, the two-point conformal blocks on a torus take just the form of a torus character \cite{Cho:2017oxl}. Therefore,  the three-point function of primary states can be given by just a shift $c \to c-1$ up to a constant factor as in \cite{KM}. However, in the case $E_A \ll E_B$, the contribution from the two-point block on a torus is nontrivial and therefore the above result is not contradict with the results in \cite{Das:2017cnv,Ku}.

\subsection{H-L-L 3pt Functions from Conformal Bootstrap}

We can also derive the behavior (\ref{asymth}) for the heavy-light-light three-point function
from a simple argument of conformal bootstrap. First note that the four-point function of  primary operators $O_1$ and $O_2$ can be expanded in terms of all states $A$ in a given 2D CFT:
\ba
\la O_1(0)O_2(x,\bar{x})O_2(1)O_1(\infty)\lb=|x|^{-E_{O_1}-E_{O_2}}\sum_A |C_{O_1O_2 A}|^2 x^{\Delta_A}\bar{x}^{\bar{\Delta}_A}, \label{fptr}
\ea
where $C_{O_1O_2 A}=\la O_1|O_2|A\lb$.
The bootstrap relation in the limit $x\to 1$ leads to the behavior
\ba
\la O_1(0)O_2(x,\bar{x})O_2(1)O_1(\infty)\lb \simeq |1-x|^{-2E_{O_2}}.
\ea
Therefore we obtain the following relation in the limit $x=\bar{x} \to 1$:
\ba
\int dE_A  D(E_A) \overline{|C_{O_1O_2 A}|^2} x^{E_A}
\sim  |1-x|^{-2E_{O_2}},
\ea
where the density of state reads $D(E_A)\sim e^{2\pi\s{\f{c E_A}{3}}}$.
It is straightforward to see that this leads to the behavior:
\be
\overline{|C_{O_1O_2 A}|^2} \sim e^{-4\pi\s{\frac{cE_A}{12}}},
\ee
which indeed reproduces (\ref{asymth}).

\subsection{Holographic CFTs}\label{sechcft}

So far our arguments assumed generic unitary CFTs in two-dimensions, where we have shown the exponential
suppressions of the off-diagonal three-point functions for the energies much larger than the central charge $E_A,E_B\gg c$.
However, if we consider the special class of CFTs, called holographic CFTs, this energy condition is relaxed as we will explain below. The holographic CFTs are characterized by the large degrees of freedom and strong interactions so that they have classical holographic duals. In two dimensional CFTs, these conditions are equivalent to the large central charge $c\gg 1$ and the sparse spectrum. Under these conditions, the Cardy formula\footnote{Note that in (\ref{cfla}) we assumed $E\gg c$ and thus we made the approximation $E-\frac{c}{12}\simeq E$. In this subsection, we choose $E$ and $c$ are both equally large and we cannot allow this approximation.} $D(E)\sim e^{4\pi\s{\frac{c}{12}\left(E-\frac{c}{12}\right)}}$
holds for any states with the energy $E=\Delta+\bar{\Delta}\geq \frac{c}{6}$ as proved in \cite{Hartman:2014oaa}, which is equivalent to a sharp confinement/deconfinement phase transition of the partition function at $\beta=2\pi$. Therefore we can employ the Cardy formula even for such relatively low energy states.

Let ask if we can apply the previous argument in section \ref{secthp}
to such states $E_{A,B}= O(c)$. The crucial point is the validity of the saddle point approximation (\ref{wwe}). Since the dominant contribution is localized at $\theta=0$, we find that the saddle point is located at $E_{A,B}-\f{c}{12}\simeq \rho=\left(\frac{2\pi}{\beta}\right)^2\cdot \frac{c}{12}$. Since we consider the low temperature phase for the right hand side of the basic relation (\ref{minv}), we have the condition $\beta<2\pi$.  From this, we obtain $E_{A,B}\geq \frac{c}{6}$. In this way, we can conclude that the exponential suppressions of the off-diagonal three-point functions (\ref{ethfa}),(\ref{eeeq}) and (\ref{ethfb}) occur in 2D holographic CFTs when $E_{A,B}\geq \frac{c}{6}$, which allow much lower energy regions compared with the
condition of the same suppressions for generic CFTs, i.e., $E_{A,B}\gg c$.

\subsection{Comparison with ETH}

The ETH (Eigenstate Thermalization Hypothesis) \cite{ETH,ETHR,ETHRR}, which we will briefly review below, is formulated for matrix elements of observables $O$ in the basis $| n \lb$ for the eigenstates of Hamiltonian $H$:
\ba
\la n|O|m \lb=f_O(E_n)\delta_{nm}+e^{-\frac{S(E)}{2}}g_{O}(E_n,E_m)R_{nm}, \label{ETH}
\ea
where $S(E)$ is the averaged entropy at the energy $E=\frac{E_n+E_m}{2}$. The matrix $R_{nm}$ is a random Hermitian matrix with zero mean and unit variance. Explicitly, we have
\be
\la\la R_{kl}R_{mn}\lb\lb=\delta_{l,m}\delta_{k,n},  \label{wick}
\ee
where $\la\la\ddd\lb\lb$ denotes the random average. The functions $f_O(E_n)$ and $g_O(E_n,E_m)$ are smooth functions of the energies $E_{n,m}$.
This behavior (\ref{ETH}) can be obtained by assuming that the eigenstates $|n\lb$ are random states, i.e., if we choose a basis $|i\lb$ such that the observable $O$ is diagonal $\la i|O|j\lb=\delta_{ij}O_i$, then the vectors $\{p^n_i\}$ defined by $|n\lb=\sum_ip^n_{i}|i\lb$ are random such that $\la\la p^{*n}_{i} p^m_j\lb\lb=\frac{1}{D}\delta_{ij}\delta_{nm}$, where $D$ is the dimension of Hilbert space.

The ETH is considered to be true in a closed quantum system with a quantum chaos.
If we assume ETH, we can show the thermalization of the observable $O$ as follows.
Consider the time evolution of expectation value $\la O(t)\lb$ for a quantum state
$|\psi\lb=\sum_{n}b_n|n\lb$:
\ba
\la O(t)\lb&=&\sum_{n,m}\la n|O|m \lb b^*_n b_m e^{i(E_n-E_m)t} \no
&=& \sum_{n}|b_n|^2 \la n|O|n \lb+\sum_{n\neq m}e^{i(E_n-E_m)t} \la n|O|m \lb b^*_n b_m .
\ea

We say that $O$ is thermalized if the (some) time average of $\la O(t)\lb$
coincides with its microcanonical prediction. For this we need to show that
(i) the time average $\bar{O}=\frac{1}{T}\int^T_0 dt\la O(t)\lb$ only depends on the energy, and it should not depend on the details of the coefficient $b_n$, and moreover, (ii) the fluctuation is very small:
$\frac{1}{T}\int^T_0 dt\la \left(O(t)\right)^2\lb-\left(\bar{O}\right)^2\sim e^{-S(E)}$. Indeed, we can confirm that (i) is satisfied because the diagonal part of
(\ref{ETH}) only depends on the energy and that (ii) is satisfied because the off-diagonal part is suppressed by the factor $e^{-S/2}$.

In the ETH, we regard $O$ as a low energy operator and take the states $|n\lb$ to be high energy states which are responsible for thermalizations. The ETH property
(\ref{ETH}) leads to the estimation when we take an appropriate average:
\be
\la\la|\la n|O|m \lb|^2\lb\lb\simeq  e^{-S(E)}(g_O(E_n,E_m))^2.
\ee
Indeed this exponentially suppressed behavior nicely agrees with our results (\ref{ethfa}), (\ref{eeeq}) and (\ref{ethfb}) for the off-diagonal part of the heavy-light-heavy three-point functions, by identifying $A=n$ and $B=m$. In this way, the modular invariance in 2D CFTs nicely reproduces the ETH property for off-diagonal three point functions. Note that for generic unitary 2D CFTs, these ETH behaviors for the averaged three point functions hold for high energy states $E_{n},E_{m}\gg c$ as we have seen in section \ref{secthp}. In particular, for holographic CFTs, these properties hold even for relatively low energy states as long as the condition $E_{n},E_{m}\geq \frac{c}{6}$ is satisfied.

Also note that the diagonal part (\ref{diagt}), which is
clearly a smooth function, gives the function $f_O(E)$ in (\ref{ETH}). For holographic 2D CFTs, we expect $f_O(E)$ is indeed 
not exponentially suppressed  because we normally have $E_\chi=O(1)$ in (\ref{diagt}) and take the large $c$ limit. For generic 2D CFTs, this ETH requirement of non-suppressed $f_O(E)$ is not guaranteed.

\subsection{Torus Multi-Point Functions and Modular Invariance}

The ETH (\ref{ETH}) argues that the off-diagonal three-point functions are not only exponentially suppressed but also are random valued. We would like to study their multi-point correlations to examine this property. For this, we consider the modular invariance of $N$-point functions on a torus $\la O(t_1)O(t_2)\ddd O(t_N)\lb$.  For simplicity we choose $t_{i}=\frac{N-i}{N}\beta$ for
$i=1,2,\ddd,N$. In the low temperature limit, this $N$-point function on a torus factorizes into
the thermal partition function times the $N$-point function on a cylinder, where the latter does not have any exponentially growing factor. As similar to the $N=2$ case (\ref{minv}), the modular invariance of the torus $N$-point function
in the limit $\beta\to 0$ leads to
\ba
\sum_{A_1,A_2,\ddd,A_N}\la A_1|O|A_2\lb \la A_2|O|A_3\lb\ddd \la A_N|O|A_1\lb\cdot e^{-\frac{\beta}{N}(E_1+E_2+\ddd +E_N)}
\sim e^{\frac{\pi^2c}{3\beta}}, \label{bnml}
\ea
where we employed the estimation of the partition function $Z(1/\beta)\sim e^{\frac{\pi^2c}{3\beta}}$. The energies $E_1,E_2,\ddd$
are those of the states $|A_1\lb,|A_2\lb,\ddd$, respectively. By using the Cardy formula as before, we finally obtain from (\ref{bnml}) the following average of the $N$ products of the three-point functions when all energies are the same $E_1=E_2=\ddd =E_N\equiv E$:
\ba
\ov{\la A_1|O|A_2\lb \la A_2|O|A_3\lb\ddd \la A_N|O|A_1\lb}\sim e^{(1-N)S(E)},  \label{avecor}
\ea
where we do not take any summations over $A_i$; also $S(E)=4\pi\s{\frac{cE}{12}}$ is the entropy for states with the energy $E$ again.

Next we would like to compare this result with that obtained by assuming the ETH behavior (\ref{ETH}), where
$R_{nm}$ is a random matrix. When $N$ is even, we find
\ba
 \ov{\la A_1|O|A_2\lb \la A_2|O|A_3\lb\ddd \la A_N|O|A_1\lb}\sim  e^{-\frac{N}{2}S(E)}\ov{\la\la R_{A_1A_2}R_{A_2A_3}\ddd R_{A_{N}A_{1}}\lb\lb} \sim e^{-(N-1)S(E)}.  \label{corewaa}
\ea
Here we evaluated the averaged random correlation function by the Wick contractions in the random average (\ref{wick}) as
\ba
\ov{\la\la R_{A_1 A_2}R_{A_2 A_3}\ddd R_{A_{N} A_{1}}\lb\lb}=\ov{\delta_{A_1A_3}\delta_{A_3A_5}\ddd\delta_{A_{N-1}A_1}+(\mbox{permutation})}
\sim e^{-\left(\frac{N}{2}-1\right)S(E)}.
\ea
Remember that the average $\ov{M}$ in the above means that for all states with the energy $E$ i.e.
$\ov{M}=e^{-NS(E)}\sum_{A_1,A_2,\ddd,A_N}M$. When $N$ is odd, we can obtain the same estimation (\ref{corewaa})
from the ETH ansatz (\ref{ETH}) by replacing one of $e^{-S(E)/2}R_{nm}$ with $\delta_{n,m}$ in the averaged random correlation function.  In this way, the CFT result (\ref{avecor}) reproduces the random matrix result (\ref{corewaa}) based on the ETH ansatz for these correlation functions. Note that again this analysis can applied to high energy states $E\gg c$ for generic 2D CFTs and to the states $E\geq \frac{c}{6}$ for holographic CFTs.

\section{Constraints from Open-Closed Duality}

Next we would like to study universal constraints for 2D CFTs with boundaries, where conformal boundary conditions are imposed. These theories are called as boundary conformal field theories. Especially we are interested in properties of the one-point functions of primary operators on a disk. For this, it is useful to employ the description in terms of boundary states, which are closed string states for boundaries.

 States which describe physical boundaries are called Cardy states $|B_\ap\lb$ \cite{Cardy:1989ir} and are given by the linear combinations of Ishibashi states $|I_k\lb$ as follows
\be
|B_\ap\lb=\sum_{k}c^\ap_k|I_k\lb,
\ee
where the label $k$ runs over those of primary states. The index $\ap$ describes the types of boundary conditions.

The Ishibashi state $|I_k\lb$ is constructed out of the primary state $|k\lb$ and its descendants.
It is given by the maximally entangled state between the left and right moving sectors:
\ba
|I_k\lb=\sum_{\vec{n}}|\vec{n},k\lb_L |\vec{n},k\lb_R.
\ea
Here the states $|\vec{n},k\lb$ make the orthonormal basis of the descendants of the form
\be
|\vec{n},k\lb=N_{\vec{n},k}\left[(L_{-1})^{n_1}\ddd (L_{-m})^{n_m}\ddd+...\right]|k\lb,
\ee
where $\{L_{m}\}$ are Virasoro generators and
$N_{\vec{n},k}$ are the overall constants to make the states orthonormal:
\be
\la \vec{n},k|\vec{m},l\lb=\delta_{k,l}\delta_{\vec{n},\vec{m}}.
\ee

\subsection{Duality for Vacuum Cylinder Amplitudes}

The open-closed duality relation for vacuum cylinder amplitudes, so called Cardy's condition \cite{Cardy:1989ir},
is written as
\be
\la B_\ap|e^{-sH_C}|B_\beta\lb=\sum_{\gamma}N^{\gamma}_{\ap\beta}\mbox{Tr}_{\gamma}
[e^{-2\pi t H_O}],  \label{oc}
\ee
where $t$ and $s$ are related by
\be
t=\frac{\pi}{s}.
\ee
Here $H_C$ and $H_O$ are the closed and open string Hamiltonians, respectively:
\ba
&& H_C=L_0+\ti{L}_0-\frac{c}{12}, \ \ \ H_O=L_0-\frac{c}{24}.
\ea
Moreover, $\gamma$ runs all primary states, and $N^\gamma_{\ap\beta}$ is a positive integer
which counts the number of open string sectors and is non-zero only when the corresponding OPE coefficient is non-vanishing.

\subsubsection{$t\to 0$ Limit}

Let us first study a familiar limit $t\to 0$. Here we assume the limit of large central charge $c\gg 1$ so that the degeneracy of primary states is approximately given by the Cardy formula (\ref{cfla}), with neglecting descendant
contributions. We write the conformal dimension as $L_0=\Delta$ in the open string sector.
In this limit, the duality condition (\ref{oc}) leads to
\ba
\bar{c}^\ap_0c^\beta_0\cdot e^{\frac{\pi c}{12t}}\sim \int d\Delta D_{open}(\Delta)e^{-2\pi t\left(\Delta-\frac{c}{24}\right)},
\ea
where $D_{open}(\Delta)$ is the density of state in open string at energy $H_O=\Delta-\frac{c}{24}$.
Here we have assumed that $c^\ap_0\neq 0$ and $c^\beta_0\neq 0$ for boundary conditions $\ap$ and $\beta$. These conditions are equivalent to those for
vacuum disk amplitudes to be non-vanishing, and they are usually satisfied as they mean the non-zero tensions of D-branes. By using the saddle point approximation formula,
\be
\int dx e^{2\pi \lambda\s{\frac{cx}{6}}}e^{-2sx}\sim e^{\frac{\pi^2 c\lambda^2 }{12s}}, \label{saddlep}
\ee
we find the following estimation for $\Delta\gg c$:
\be
D_{open}(\Delta)\sim e^{2\pi \s{\frac{c\Delta}{6}}},
\ee
which gives the Cardy formula for open strings.

\subsubsection{$s\to 0$ Limit}

Now let us turn to the less familiar limit $s\to 0$.
First we assume $\ap=\beta$. We define the chiral conformal dimension as $L_0=\ti{L}_0=\Delta_k$ for the primary state $| k \lb$.
In this case we get from (\ref{oc}):
\ba
\int d\Delta_k \overline{|c^\ap_k|^2}  e^{-(2\Delta_k-\frac{c}{12})s}D_{closed}(\Delta_k)\sim e^{\frac{\pi^2 c}{12s}}.
\ea
Here the density of state (note that $L_0=\ti{L}_0$ is imposed for boundary states) is given by the same as the chiral or open string Cardy formula for $\Delta_k \gg c$:
\be
D_{closed}(\Delta_k)\sim e^{2\pi \s{\frac{c\Delta_k}{6}}},
\ee
which is a square root of the full Cardy formula (\ref{cfla}).
This leads to the following estimation
\be
\overline{|c^\ap_k|^2}\sim 1,  \label{pppqwe}
\ee
where an appropriate average over $k$ with fixed dimension $\Delta_k$ is taken.

Now we take the different boundary conditions $\ap\neq \beta$.
In this case, we have $N^0_{\ap\beta}=0$ and thus there should be a gap in the open string spectrum.
We call the minimum of the conformal dimension for the open string in the channel $\gamma$ as $\Delta^{min}_\gamma$.
Then, the relation (\ref{oc}) leads to
\ba
\int d\Delta_k \overline{\bar{c}^\ap_k  c^\beta_k}  e^{-\pa{2\Delta_k-\frac{c}{12}}s}D_{closed}(\Delta_k)\sim e^{\frac{\pi^2 (c-24\Delta^{min}_\gamma)}{12s}}.
\ea
Thus we obtain the estimation
\ba
\overline{\bar{c}^\ap_k  c^\beta_k} \sim
e^{2\pi\s{\frac{(c-24\Delta^{min}_\gamma)\Delta_k}{6}}
-2\pi\s{\frac{c}{6}\Delta_k}}=e^{-\frac{\pi c}{3}\left(1-\s{1-\frac{24}{c}\Delta^{min}_\gamma}\right)\s{\frac{6}{c}\Delta_k}}.  \label{corap}
\ea
This can be understood as the correlations between the coefficients (or disk one-point functions) $c^\ap_k$ and $c^\beta_k$. Compared with (\ref{pppqwe}), we find that the correlation is reduced as the gap in the open strings between the two boundaries $\ap$ and $\beta$ develops.

\subsubsection{Holographic Interpretation}

Here we would like to give a holographic interpretation of the result (\ref{corap}), which is motivated by the argument in \cite{KM} for the diagonal three-point functions.%
\footnote{A gravity dual interpretation is expected to be possible only for a holographic CFT with the condition of the large central charge and the sparse spectrum. However, as we will show below, we can reproduce the CFT result (\ref{corap}) without using the condition for the current case analogously for the Cardy formula (\ref{cfla}).}
We first consider the case with $1 \ll \Delta^{min}_\gamma \ll \frac{c}{24}$, where the expression of (\ref{corap}) reduces to
\begin{align}
\overline{\bar{c}_k^\alpha c_k^\beta}  \sim e^{- 2 \pi \Delta^{min}_\gamma  \sqrt{ \frac{24 }{c}\Delta_k} } .
\label{cclargec}
\end{align}
We then include the corrections as in (\ref{corap}) for $\Delta^{min}_\gamma < \frac{c}{24}$.

\begin{figure}
	\centering
	\includegraphics[width=7cm]{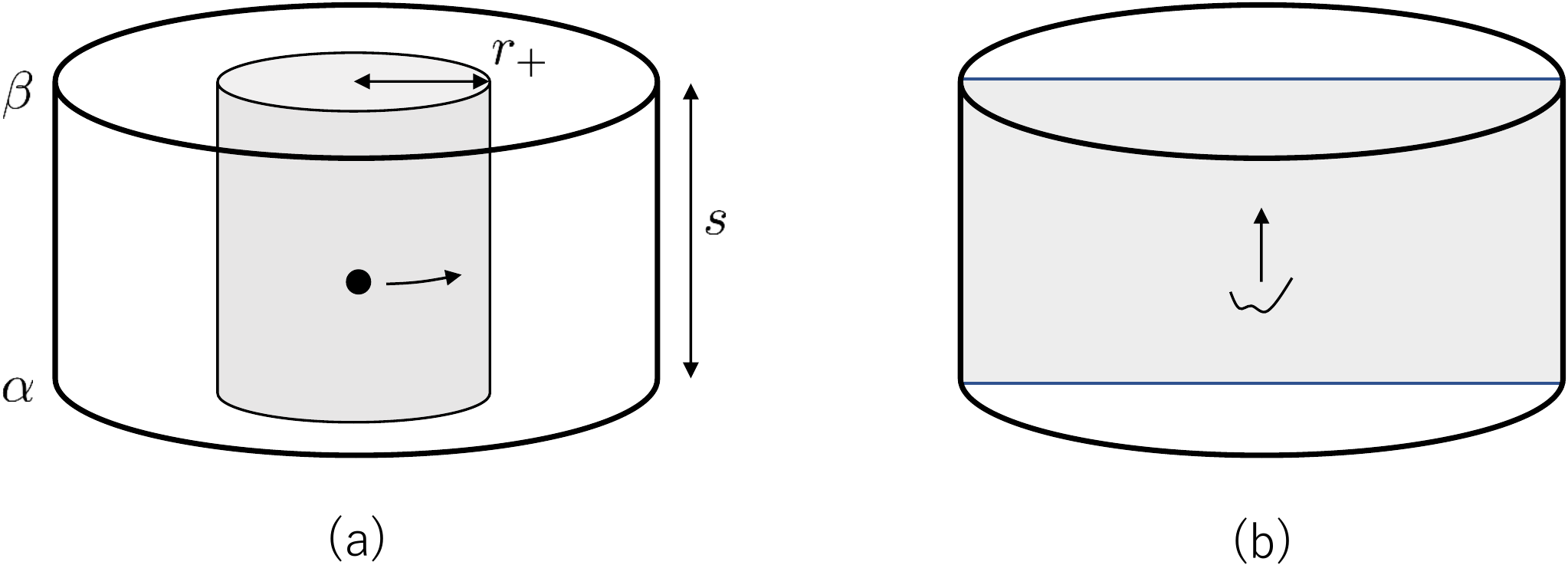}
	\caption{(a) A closed string state at high energy and a light open string state are approximately described by the BTZ black hole and a perturbative scalar particle, respectively.
	(b) A bulk light particle dual to a boundary open string state can be regarded as a low energy excitation of a bulk open string on a brane.}
	\label{Gravity}
\end{figure}

We are interested in the exchange of closed strings with high energy $ E_k = 2 \Delta_k \gg c$ between the boundary states.
The high energy state of closed string is dual to the BTZ black hole, whose horizon area is
\begin{align}
A = 2 \pi r_+  , \quad
r_+ = \sqrt{\frac{24 \Delta_k}{c} - 1}  .
\label{area}
\end{align}
For $\Delta^{min}_\gamma \ll \frac{c}{24}$,
the open string state with energy $\Delta^{min}_\gamma$ can be described as a perturbative scalar particle with mass $m^{min}_\gamma \sim \Delta^{min}_\gamma$.
The leading order contribution comes from the scalar particle running around the BTZ black hole, see figure \ref{Gravity}(a).

The other assumption $\Delta^{min}_\gamma \gg 1$ implies that the contribution can be evaluated by the geodesic wrapping the black hole horizon as $\exp \left(-m_\gamma^{min} A \right) $.
Using  $m^{min}_\gamma \sim \Delta^{min}_\gamma$ and the area of horizon \eqref{area} we reproduce \eqref{cclargec} for $\Delta_k \gg c$.

Next we relax the condition of $\Delta^{min}_\gamma$ as $\Delta^{min}_\gamma < \frac{c}{24}$.
In this case, we should take care of the back-reaction of the particle since it would create a conical defect geometry. Let us consider a particle $\phi$ with mass $m_\phi$ and its energy $E_\phi$ evaluated at the boundary of AdS. The relation between them is given by (see (27) of \cite{KM})
\begin{align}
m_\phi = \frac{c}{6} \left( 1 - \sqrt{1 - \frac{12 E_\phi}{c}}\right) .
\label{conical}
\end{align}
However, we should be careful for applying the formula to our setup since we are dealing with a bulk particle dual to a boundary open string state. The bulk particle is regarded as a low energy excitation of a bulk open string attached to a brane as in figure \ref{Gravity}(b), see, e.g., \cite{Takayanagi:2011zk,Fujita:2011fp} for AdS/BCFT.
Since a pair of open strings create a closed string, we should set $ 2 m_\gamma^{min} = m_\phi $ and $2 \Delta_\gamma^{min} = E_\phi $. Therefore, we have
\begin{align}
m_\gamma^{min}  = \frac{c}{12} \left( 1 - \sqrt{1 - \frac{24 \Delta_\gamma^{min}}{c}} \right) \, .
\end{align}
With this expression of $m_\gamma^{min}$ and the area of horizon \eqref{area},
the contribution from the geodesic of the particle,
$\exp \left(-m_\gamma^{min} A \right) $, reproduces (\ref{corap})  for $\Delta_k \gg c$.

\subsection{Duality for 1pt Function on Cylinder}

Finally we would like to study the open-closed duality for a
cylinder one-point function of a primary operator $O$:
\ba
\la B_\ap|e^{-s_1H_C}Oe^{-s_2H_C}|B_\beta\lb=\sum_{\gamma}N^{\gamma}_{\ap\beta}
\mbox{Tr}_{\gamma}[Oe^{-2\pi t H_O}],  \label{ooc}
\ea
where $t=\frac{\pi}{s_1+s_2}$. Again we assume the limit of large central charge $c\gg 1$.

In the limit $s_1,s_2\to 0$, this relation is expressed as follows:
\ba
&& \int d\Delta_k d\Delta_l e^{2\pi\s{\frac{c\Delta_k}{6}}}e^{2\pi\s{\frac{c\Delta_l}{6}}}
\overline{\la k|O|l\lb} e^{-(2\Delta_k-c/12)s_1} e^{-(2\Delta_l-c/12)s_2}
\bar{c}^\ap_{k} c^{\beta}_{l} \no
&& \simeq N^\gamma_{\ap\beta}\la \gamma|O|\gamma\lb_{open} \cdot e^{\frac{\pi^2}{12(s_1+s_2)}(c-24\Delta^{min}_{\gamma O})}.\label{relw}
\ea
Where $\gamma$ is the open string state with the smallest conformal dimension
$L_0=\Delta^{min}_{\gamma O}$, which satisfies
$N^\gamma_{\ap\beta}\neq 0$ and $\la\gamma|O|\gamma\lb_{open}\neq 0$. Note that by definition we have
\be
\Delta^{min}_{\gamma O}\geq \Delta^{min}_{\gamma}.  \label{vvv}
\ee

First we evaluate contributions from the diagonal part.
We can employ the known formula (\ref{diagt})  for the diagonal parts of three-point functions (with  $L_0=\bar{L}_0=\Delta=E/2$). Then, by using the saddle point formula and the previous formula (\ref{corap}), we can estimate the relation (\ref{relw}) and obtain
the following relation:
\ba
\s{1-\frac{24}{c}\Delta^{min}_{\gamma O}}\geq \s{1-\frac{24}{c}\Delta^{min}_\gamma}-2\left(1-\s{1-\frac{24}{c}\Delta_\chi}\right).
\label{kwrb}
\ea
The inequality should be saturated if the diagonal part gives the dominant contributions.
Note that the inequality is consistent with (\ref{vvv}).

 Next let us estimate contributions from the off-diagonal parts of three-point functions. The open-closed duality tells us the relation
\be
\sum_{k,l} \overline{\la k|O|l\lb \bar{c}^\ap_k c^\beta_l} e^{-2\Delta_k s_1}e^{-2\Delta_l s_2}
\sim e^{\frac{\pi^2}{12(s_1+s_2)}(c-24\Delta^{min}_{\gamma O})}. \label{xxxt}
\ee
In particular, if we choose $s_1=s_2=s/2$, we find
\ba
\int d\Delta_k d\Delta_l e^{2\pi\s{\frac{c}{6}\Delta_k}+2\pi\s{\frac{c}{6}\Delta_l}}\cdot
e^{-(\Delta_k+\Delta_l)s}\cdot \overline{\la k|O|l\lb \bar{c}^\ap_k c^\beta_l}\sim e^{\frac{\pi^2}{12s}(c-24\Delta^{min}_{\gamma O})}.
\ea
When $\Delta_k\simeq \Delta_l$, this relation leads to the behavior:
\be
\overline{\la k|O|l\lb \bar{c}^\ap_k c^\beta_l} \lesssim
 e^{-\pi\left(2-\s{1-24\Delta^{min}_{\gamma O}/c}\right)\left(\s{\frac{c}{6}\Delta_k}+\s{\frac{c}{6}\Delta_l}\right)},  \label{ocxx}
\ee
where the inequality is saturated when the off-diagonal contributions are dominant. In other words, either (\ref{kwrb}) or
(\ref{ocxx}) should be saturated in order for the open-closed duality relation (\ref{ooc}) to be satisfied.

\section{Conclusions}

In this article, we studied implications of modular dualities in 2D CFTs. In particular, we analyzed the modular invariance of two-point functions on a torus and the open-closed duality of cylinder amplitudes with the input of density of high energy states given by the Cardy's formula.

The modular invariance of two-point functions leads to non-trivial constraints
on the behavior of three-point functions where two of the operators are heavy, whose dimensions are much larger than the central charge, and the other is much lighter than them. We found that the off-diagonal part of three-point functions, for which the case with two different heavy states, is exponentially suppressed by the entropy as $\sim e^{-\frac{S(E)}{2}}$ under an appropriate average. Interestingly, this non-trivially satisfies the important condition required for the ETH. This implies that high energy states in generic 2D CFTs have the crucial property necessary for thermalizations. To see more details how much a given 2D CFT is chaotic, we need to understand the properties of the matrix $R_{nm}$ in (\ref{ETH}). For truly chaotic CFTs, we expect that $R_{nm}$ becomes a random matrix. Even though we leave full studies of the random matrix property in 2D CFTs for a future problem, we found some evidence for this by studying a class of multi-point correlations of the off-diagonal three-point functions.

It is natural to expect that for integrable CFTs, only with particular choices of $n$ and $m$, this matrix takes non-trivial values. In such an analysis, we expect that large central charge CFTs with large spectrum gaps, namely holographic CFTs, will play an important role \cite{Hartman:2014oaa}. We found that for 2D holographic CFTs, both the exponential suppressions
and the expected behavior for the multi-point correlations of the three-point functions, occur even for relatively low energy states. These reinforce the chaotic properties for holographic CFTs. It will be an intriguing future problem to examine more closely the randomness of the matrix $R_{mn}$ for various 2D CFTs. The constraints for three-point functions also come from the conformal bootstrap. Recently, an interesting transition phenomenon for conformal blocks, depending on whether the conformal dimensions are larger than $c/32$ or not, was observed in \cite{KuTa,Ku}. It would be interesting to consider its implication in terms of the behavior of three-point functions.

The open-closed duality for vacuum cylinder amplitudes turned out to predict interesting behaviors of disk one-point functions (or equally the coefficients of boundary states) for various conformal boundary conditions. We found that as the mass gap in open stings between two boundaries gets larger, the correlation between the two disk one-point functions for the boundary conditions is reduced. We gave a holographic explanation for this result. We also studied implications of the open-closed duality for cylinder one-point functions. This leads to interesting constraints on the average of a three-point function times
two disk one-point functions. It would be interesting to look at explicit examples of boundary states in solvable CFTs, such as various orbifold CFTs, to see how the chaos is
related to these properties of boundary states.

\subsection*{Acknowledgements}
 We thank Tokiro Numasawa very much for useful discussions.
 We are also grateful to Kanato Goto and Chen-Te Ma for conversations.
 TT would also like to thank Stanford Institute for Theoretical Physics for their hospitality where this work was completed. YH and TT are supported by JSPS Grant-in-Aid for Scientific Research (A) No.16H02182. YK is supported by JSPS fellowship. TT is also supported by the Simons Foundation through the ``It from Qubit'' collaboration and by World Premier International Research Center Initiative (WPI Initiative) from the Japan Ministry of Education, Culture, Sports, Science and Technology (MEXT).


\appendix

\end{document}